\documentclass{article}
\usepackage{lscape}

\usepackage{graphicx,pstricks}

\title{\bf  Reassessing Accuracy Rates of Median Decisions }

\author{\bf Andrea Capotorti, \ \ \ \ \ \ \ \ \ \ \ \ Frank
Lad \ \ \ \ \ \ \ \ and \ \ \ \ Giuseppe Sanfilippo  \\
University of Perugia \ \ \ \ \ University of
Canterbury \ \ \ \ \ \ \ \ \ \ \ \ \  University of Palermo}

\date{  }

\begin{document}

\maketitle


\begin{abstract}
We show how Bruno de Finetti's fundamental theorem of prevision has
computable applications in statistical problems that involve only
partial information.  Specifically, we assess accuracy rates for
median decision procedures used in the radiological diagnosis of
asbestosis. Conditional exchangeability of individual radiologists'
diagnoses is recognized as more appropriate than independence which
is commonly presumed. The FTP yields coherent bounds on
probabilities of interest when available information is insufficient
to determine a complete distribution. Further assertions that are
natural to the problem motivate a partial ordering of conditional
probabilities, extending the computation from a linear to a
quadratic programming problem.
\end{abstract}

\section{Introduction}

At the invitation of Maurice Fr\'{e}chet, Bruno de Finetti (1937)
delivered six lectures to the Institute Henri Poincar\'{e} in Paris.
In them he showed how probability, and more generally expectation,
can be defined in terms of a price, and how the concept of coherence
of an array of such prices assessed by an individual generates a
unified theory of probability and expectation. He termed the
asserted values for observable quantities as ``previsions,''  and
introduced the notation of P(E) and P(X) as common to both
assertions.  The lectures also introduced the judgment of
exchangeability as representing symmetry in one's uncertain
attitudes about a sequence of quantities, implying an inferential 
procedure for learning
about future events in the sequence from the observation of earlier
events.  The presentation formalized what has been called ``de
Finetti's representation theorem'' for exchangeable distributions:
that infinitely extendible exchangeable assessments of a sequence of
events can be represented as mixtures of conditionally independent
distributions. Finally, he explained what he later termed ``the
fundamental theorem of probability'' (de Finetti, 1974, Section
3.10) showing how the principle of coherency determines precise and
computable bounds on the probability for {\it any event} if that
probability is to cohere with a list of
probabilities and conditional probabilities already specified. \\

The statistical community today is still coming to grips with the
implications of these lectures.  They provide a completely
different foundation for the prospects and limitations of
statistical inference than was the framework in which mainstream
statistical theory and practice developed during the twentieth
century.  The most widely
known of de Finetti's results is his representation theorem for
exchangeable distributions.  Interpreted as a formulation for how
independent random variables can be used for transforming a prior
distribution for unknown probabilities into a posterior
distribution, it is honored by many proponents as a cornerstone for
the current practice of Bayesian statistics.  Unknown to many
statisticians who are familiar with Bayesian computational methods,
this is not at all the way de Finetti thought of his mathematical
constructions.  A sophisticated yet practical 
introduction to de Finetti's outlook
and statistical methodology appears in the text of Lad (1996). \\

In the present article, we address an important statistical problem
involving multiple radiologists assessing the condition of
asbestosis in lung tissue by means of X-rays. In doing so, we
introduce the basic meaning and relevance of the judgment
to regard a sequence of quantities exchangeably, and we show how the
fundamental theorem of probability can be used directly in this
application to yield computable interval probabilities for the accuracy
rates of median diagnoses. \\

In Section 2 we present summary substantive background for 
understanding the problem of asbestosis diagnosis.  We then present a
brief introduction to the meaning of exchangeability in Section 3,
and to the computational use of de Finetti's fundamental theorem of
probability in Section 4.  In Section 5 we show how the judgment of
conditional exchangeability regarding the assessments of three
radiologists, along with some other appropriate judgments, provides
inputs for the computation of probability bounds
for accuracy rates of ``median diagnostic procedures.''
Computational results are discussed in Section 6. Throughout the
article we use the mathematical syntax and language of de Finetti's
operational-subjective construction of probability and statistical
method.  The {\it TAS} on-line repository contains supplements to the 
content of Sections 4,5 and 6.  More extensive discussion and
literature references on all matters can be found in the research report
of Capotorti, Lad and Sanfilippo (2003).

\section{Accuracy rates of asbestosis diagnosis using three radiologists}

The condition of asbestosis (fibrosis of the lung) can be identified
precisely only by removing some tissue from a lung and examining it
for metallic nodules using histological laboratory procedures.  This
is considered to be the gold-standard of asbestosis detection.
Because there is not much that current medical therapy can do for a
patient who has this condition, this histological procedure is
seldom undertaken except during autopsies of patients who have died
of lung cancer.  A cheaper and less invasive, but less precise
method of diagnosis is typically followed using lung X-rays.
Asbestosis may exhibit itself on an X-ray by a shadowy character to
the film. Assessment is so difficult that radiologists must be
specially trained to achieve the qualification of a ``B-reader'' to
be permitted to read them.  There are graded categories of severity
of the condition that can be assessed according to the density of
the shadow. Official standards of the International Labor
Organization require that at least two of three specialized film
readers (each blinded to the assessment of the others) must assess
the film in a category ``at least as bad as a specified standard''
in order for the subject to be recognized as having asbestosis. Such
a ``median diagnosis'' is required in legal proceedings for the
award of damages to a worker to be paid by an employer.  \\

    We shall designate by F the event that
a subject actually has fibrosis of the lung, which could be detected
by histological examination.  The measurement $F = 1$ denotes the
presence of asbestosis detected by such an exam, while $F = 0$
denotes no presence.  The decisions of the individual radiologists
to assign the X-ray to the category of ``asbestosis at least as bad
as the minimal standard'' are denoted as the events $D_i$ for i = 1,
2, and 3. Each $D_i = 1$ if the i$^{th}$ radiologist makes a
positive diagnosis, while $D_i = 0$ if the diagnosis is negative.
Throughout this article we use tilde notation to denote the negation
of an event. For example, a negative diagnosis can also be denoted
by $\tilde{D_i}$.  The median decision, denoted by $D^*$,
is the event that the sum of the individual diagnosis events is
at least 2, i.e., $D^* \equiv (\Sigma_{i=1}^3 D_i \geq 2)$. \\

In any diagnostic problem, there are four conditional probabilities
that characterize the accuracy of a physician's expected diagnostic
performance.  Specified in terms of an individual radiologist in
this problem, these are the probabilities $P(D_i|F)$,
$P(\widetilde{D}_i|\tilde{F})$, $P(F|D_i)$, and
$P(\tilde{F}|\widetilde{D}_i)$. Standard terminology refers to these 
probabilities 
as the sensitivity, specificity, positive predictive value and
negative predictive value of a diagnosis.  Ideally, all four of
these probabilities would be large, as close as possible to 1 in
each case.  The goal of our analysis is to identify the 
difference between these
characteristic diagnosis probabilities for individual B-readers and
the corresponding diagnosis probabilities for the median decision
procedure. These are denoted by $P(D^*|F)$,
$P(\widetilde{D}^*|\tilde{F})$, $P(F|D^*)$, and
$P(\tilde{F}|\widetilde{D}^*)$. The hope and expectation is that
each of these probabilities would exceed the corresponding accuracy
rate of an individual reader's diagnosis. In the article that
originally brought this problem to our attention, Tweedie and
Mengersen (1999) estimated the median decision accuracy rates using
the assumption that the decisions of the three radiologists are
conditionally independent given F and also 
given $\tilde{F}$. For reasons that we shall now explain, we 
propose that the judgment of ``conditional exchangeability'' is more
appropriate to assessments of
the mutually blind diagnosis decisions by the three radiologists. 

\section{Regarding events exchangeably}

The stochastic independence of three events is commonly defined by
the condition that the probability of joint occurrence of any
two or three of them equals the product of their marginal
probabilities. This implies that $P(E_i|E_j) = P(E_i)$
and $P(E_i|E_jE_k) = P(E_i)$ for any substitution of $i, j$ and $k$
by $1$, $2$ or $3$. Motivation for the application of independence
is proposed as the lack of any causal relation among the 
events. This interpretation of independence originated within a
conception of probabilities as objective characteristics of nature
that can only be estimated.  In this context, statistical analysis
is typically conducted while making ``assumptions'' about
probabilities, such as the independence
of relevant events. \\

Consider, for example, the decisions of three radiologists
concerning the exhibition of asbestosis on an X-ray.  Because the
diagnostic judgments of any two of the radiologists are unknown to
the third, they could not possibly have had any causal influence on
the judgment of the third. According to common conception then, the
three diagnosis decisions might well be considered to be
independent, and even conditionally independent given F and given
$\tilde{F}$.  The state of the X-ray would be considered to cause an
individual to make the diagnosis $D_i$ or $\tilde{D_i}$, at least
probabilistically, not the diagnosis of the
other two readers. \\

Bruno de Finetti insisted that probabilities are not
unobservable properties of nature, but rather representations of
individuals' uncertain assertions about the observable facts of
nature. Thus, when considering the difference between, say, an
assertion of $P(E_2)$ and an assertion of $P(E_2|E_1)$ we are not
considering the effect of $E_1$ on $E_2$, but rather of the
information that {\it conditional knowledge} of $E_1$ would have
on {\it one's uncertain assessment} of $E_2$. The paradigmatic 
applications of the concept of ``stochastic
independence'' to statistical analysis in objectivist thinking
involve ``random experiments'' conducted under ``identical
conditions.''  When probabilities are recognized as the
representations of individuals' uncertainties about events rather
than as properties of the events, it is evident that experimental
observations of this type are {\it not} regarded independently. Why
do we conduct experiments in the first place?  We design and conduct
them because we are uncertain what is going to happen, and because
we would like to learn about what may happen in the later
experiments in the sequence from what we observe about the earlier
ones.  We typically expect to assert different values for $P(E_2)$,
for $P(E_2|E_1)$ and for $P(E_2|\widetilde{E_1})$;  and similarly we
might assert different values for $P(E_3)$ and for $P(E_3|E_1E_2)$,
$P(E_3|E_1\widetilde{E_2})$, $P(E_3|\widetilde{E_1}E_2)$ and
$P(E_3|\widetilde{E_1}\widetilde{E_2})$. An interesting condition
among these is that $P(E_3|E_1\widetilde{E_2})$ may well equal
$P(E_3|\widetilde{E_1}E_2)$.  We shall pursue this condition
further. \\

One feature of opinions that is common to assessors of such
experiments is that {\it the order in which observed successes and
failures arrive} is regarded as irrelevant to opinions
about subsequent results, even among people who may dispute how
likely a success may be.  It is because the different experiments
are conducted in the same way every time that we do not especially
expect the successes to come early in the sequence, late in the
sequence, or especially alternating.  This is the feature 
that de Finetti characterized as the judgment to regard a
sequence of events exchangeably:\\
\noindent {\bf Definition:} \ A sequence of N events is regarded
exchangeably if the probability for any particular string involving
K successes and (N-K) failures is assessed identically, no matter
what the order in which the successes and the failures
arrive.  This must be true for each value of K between 1 and
(N-1).$ \ \ \ \ \ \diamond$ \\
Agreement to regard the order of successes as
irrelevant implies that disagreements among disputants can be
reduced by coherent inference from the results of experiments. Three
features of exchangeable distributions are especially
worth noting.\\

The assessed probability for any
particular ordered string of successes and failures depends only on one's
probability that the sum of the successes equals the sum in that string.
Specifically, if N events are regarded exchangeably, then for any permutation
of the subscripts on the events denoted by E's in the following
expression, we require the identity
\begin{center}
$P(E_1E_2...E_K\tilde{E}_{K+1}\tilde{E}_{K+2}...\tilde{E}_N) \ =
\ P(S_N = K)\:/\:^NC_K$ \ \ ,
\end{center}
where $S_N$ denotes the sum of the N events.
For there are $^NC_K$ distinct ways to permute
the subscripts and yield a distinct sequence of successes and
failures. \\

If events in a sequence are regarded independently and with identical
probabilities (iid), then the sequence is also
regarded exchangeably.  For in this case, designating the common value
of each $P(E_i)$ by $\theta$, \
$P(E_1E_2...E_K\tilde{E}_{K+1}\tilde{E}_{K+2}...\tilde{E}_N) \ =
\ \theta^K(1-\theta)^{(N-K)}$ for any permutation of the
subscripts.  Thus iid distributions over a sequence of events are
exchangeable distributions.  However, exchangeable distributions are not
necessarily iid. De Finetti's representation theorem identifies the precise
relation between iid distributions and exchangeable distributions.  An
enjoyable elementary exposition of this theorem appeared
in the article of Heath and Sudderth (1976):  \\
\noindent
{\bf Theorem:} \ If a distribution for N events is exchangeable
and can be extended to a distribution over any larger number of events
as an exchangeable distribution, then for any value of K between 1 and
(N-1) and for any permutation of the subscripts on the E's,
\begin{center}
$P(E_1E_2...E_K\tilde{E}_{K+1}\tilde{E}_{K+2}...\tilde{E}_N) \ =
\ \int_0^1\ \theta^K(1-\theta)^{(N-K)} \ dF(\theta) \ \ \ , \ \ \ $
\end{center}
for some mixing distribution function $F(\theta)$.  $\ \ \ \ \ \ \ \ \
\ \ \ \ \ \ \ \ \ \ \ \ \ \ \ \ \ \ \ \ \ \ \ \ \ \ \ \ \ \ \ 
\ \ \ \ \ \ \ \ \ \ \ \ \ \ \ \ \ \ \ \ \ \ \ \ \ \ \ \ \ \ \diamond$ \\

A misinterpretation of this theorem proposes it
as supporting a procedure for ``updating a prior distribution'' for the
``true probability'' of iid events to a posterior distribution. To 
the contrary, the usefulness of the theorem is in providing a
computational method for sequential forecasting procedures based on
the formula it implies for
$P(E_{N+1}|E_1E_2...E_K\tilde{E}_{K+1}\tilde{E}_{K+2}...\tilde{E}_N)$
for any value of K and for any permutation of the subscripts.  Clearly,
the events are {\it not} iid.  See Lad (1996, Sections 3.8 - 3.12).\\

The judgment of exchangeability has direct relevance to assessments 
of X-rays made by three experts.
No one is sure whether an expert reading an X-ray will conclude with a
diagnosis $D=1$ or $D=0$.  B-readers'
success rates for diagnosing $D=1$ when in fact $F=1$, and in diagnosing
$D=0$ when $F=0$ are both unknown;  and experts may disagree in their
uncertainties about these rates.
However, it is widely agreed that uncertain assertions about successful
diagnoses must satisfy permutation properties such as
\begin{equation}
\begin{array}{ccccl}
P(D_1\widetilde{D}_2\widetilde{D}_3|F) \ & = & \
P(\widetilde{D}_1D_2\widetilde{D}_3|F)
\ & = & \ P(\widetilde{D}_1\widetilde{D}_2D_3|F), \ \ \ \ {\rm and} \\
P(D_1D_2\widetilde{D}_3|F) \ & = & \ P(D_1\widetilde{D}_2D_3|F) \ & = & \
P(\widetilde{D}_1D_2D_3|F)\ .  \ \ \ \ \
\end{array} \label{eq:condexcheqn}
\end{equation}
Because the three B-readers are regarded as otherwise indistinguishable 
experts, it is considered that if the patient
actually has asbestosis (the condition F) it is just as likely that
any one of them makes a diagnosis that dissents from the other two.
This is considered true both when one or two of the three make a
positive diagnosis. 
All together, these four equalities represent the judgment of 
conditional exchangeability about the experts' diagnoses given $F$.  
This is a very important distinction from the judgment
of conditional independence, because it recognizes explicitly that the
diagnosis by any expert would be informative about the likely
diagnoses by the others.  Similar equalities of conditional
probabilities
would pertain when the conditioning event is $\tilde{F}$ as well.

\section{The fundamental theorem of probability and its extensions }

The fundamental theorem of probability was first described in
nugatory form in de Finetti's Paris lectures, but was named ``the
fundamental theorem'' only in his swan-song text, translated into
English in 1974.  We describe its working here in a two-part 
numerical example, which also illustrates how the assertion of 
exchangeability can be relevant to the solution of a problem.  The issue
of accuracy rates of median diagnosis procedures will then provide a
real application of the use of the theorem in its most extended form. \\

Consider two logically independent events, $E_1$ and $E_2$.  Logical
independence means that it is possible that either, both or neither
of these events can occur.  Let $E_3$ be the
event that occurs only if $E_1 = E_2$. In de Finetti-style  notation
which recognizes events as numbers, this event is determined
arithmetically via  $E_3 = 1 + 2E_1E_2 - E_1 - E_2$. 
Now suppose
that information is available to motivate the probability assertions
$P(E_1) = .7$ and $P(E_2) = .2$.  What possible values may be
asserted for $P(E_3)$ if this probability is to cohere with the two
given probabilities? The computational procedure specified by the
FTP for yielding the
solution to this problem proceeds as follows, in four steps:\\

\noindent
{\bf 1.}  Define a column vector of the quantities that are involved in the
problem, beginning with the quantities whose prevision is specified as
``given'' in the problem, and ending with the quantity whose unspecified
prevision is under consideration.  In this example, this would be the
vector {\bf E}$_3 \equiv (E_1, E_2, E_3)^T$. \\

\noindent
{\bf 2.}  Make a matrix whose columns list {\it all} the possible
observable values of the quantity vector specified in step {\bf 1}.
This matrix is called the {\it realm} of that vector.  Here
\begin{equation}
{\bf R} \; \left( \begin{array}{c}
E_1\\
E_2\\
E_3 \equiv (E_1 = E_2)
\end{array} \right) \; \; \;  =  \; \; \;
\left( \begin{array}{c}
\ 0 \ \ 0 \ \ 1 \ \ 1\ \\
\ 0 \ \ 1 \ \ 0 \ \ 1\ \\
\ 1 \ \ 0 \ \ 0 \ \ 1\
\end{array} \right)  \label{eq:realmeqn}
\end{equation}

\noindent {\bf 3.} Realize that each of the quantities in the vector
defined by step {\bf 1} can be expressed as a linear combination of
four events that identify the columns of the realm matrix.  This
would be the vector $(\widetilde{E}_1\widetilde{E}_2,
\widetilde{E}_1E_2, E_1\widetilde{E}_2, E_1E_2)^T$.  The
coefficients for these linear combinations are specified in the
corresponding rows of the realm matrix defined in equation
(~\ref{eq:realmeqn}). Although probabilities for these four
events are {\it not given} as conditions for the problem, we know
that these probabilities must sum to 1 because these events
constitute an exclusive and exhaustive partition. Moreover, since
the prevision (expectation) of any linear combination of events must
equal the same linear combination of the probabilities for those
events, and since the probabilities for the first two components of
the vector {\bf E}$_3$ {\it are given} in this problem, we have
three linear conditions on the four partition probabilities {\bf
q}$_4 \ = \ (P(\widetilde{E}_1\widetilde{E}_2), \
P(\widetilde{E}_1E_2), P(E_1\widetilde{E}_2, E_1E_2))^T$. We
can express them by the matrix equation
\begin{equation}
P \, \left( \begin{array}{c}
\ E_1\ \\
\ E_2\ \\
\ \ 1 \
\end{array} \right) \; \; \;  =  \; \; \;
\left( \begin{array}{c}
\ 0 \ \ 0 \ \ 1 \ \ 1\ \\
\ 0 \ \ 1 \ \ 0 \ \ 1\ \\
\ 1 \ \ 1 \ \ 1 \ \ 1\
\end{array} \right) \ {\bf q}_4 \; \; \; = \; \; \;
 \left( \begin{array}{c}
\ .7\ \\
\ .2\ \\
\ 1 \
\end{array} \right) \label{eq:probconsteqn}
\end{equation}
\noindent {\bf 4.}  Finally, since equation (\ref{eq:realmeqn})
shows that $E_3$ is also a linear function of the
partition events, with 
linear coefficients specified by the third row of the realm matrix, it
must also be true that $P(E_3)$ equals this linear
combination of the incompletely specified vector {\bf q}$_4$, viz.,
$P(E_3) = ( \ 1 \ 0 \ 0 \ 1 \ )${\bf q}$_4$. \\

The four steps of this procedure determine bounds for the
probability assertion $P(E_3)$ that would cohere with the conditions
given in this problem.  They can be computed via two linear
programming problems: \ Find the vectors {\bf q}$_4^*(min)$ and {\bf
q}$_4^*(max)$ that minimize and maximize $( \ 1 \ 0 \ 0 \ 1 \ )${\bf
q}$_4$, respectively, subject to the three linear conditions on {\bf
q}$_4$ displayed in equation (\ref{eq:probconsteqn}).  The numerical
solution for the minimum coherent value of $P(E_3)$ is .10,
corresponding to the vector {\bf q}$_4^*(min) \ = \ (.1, .2, .7,
0)^T\,$. The maximum value for $P(E_3)$ is
.50, corresponding to {\bf q}$_4^*(max) \ = \ (.3,0,.5,.2)^T\,$.\\

Figure 1 displays the column vectors composing the realm matrix defined 
in equation (\ref{eq:realmeqn}), each represented by a bold point.  
The polyhedron that connects them is called their convex hull.  
A coherent prevision for the vector of
unknown quantities $(E_1, E_2, E_3)^T$ must be expressible as some
convex combination of these four vertices.  Geometrically, this means 
that the prevision vector must be an interior point or a boundary point 
of the convex hull. The
specification of the first two components of $P(E_1, E_2, E_3)^T$ as
$.7$ and $.2$ means further that the vector of all three prevision values
must lie on the dashed line segment that touches two edges of the
hull in Figure 1. The extreme possibilities for $P(E_3)$ correspond
to points on the ends of this line segment.
\begin{figure}[htbp]
\begin{center}
	\includegraphics[width=0.8\linewidth]{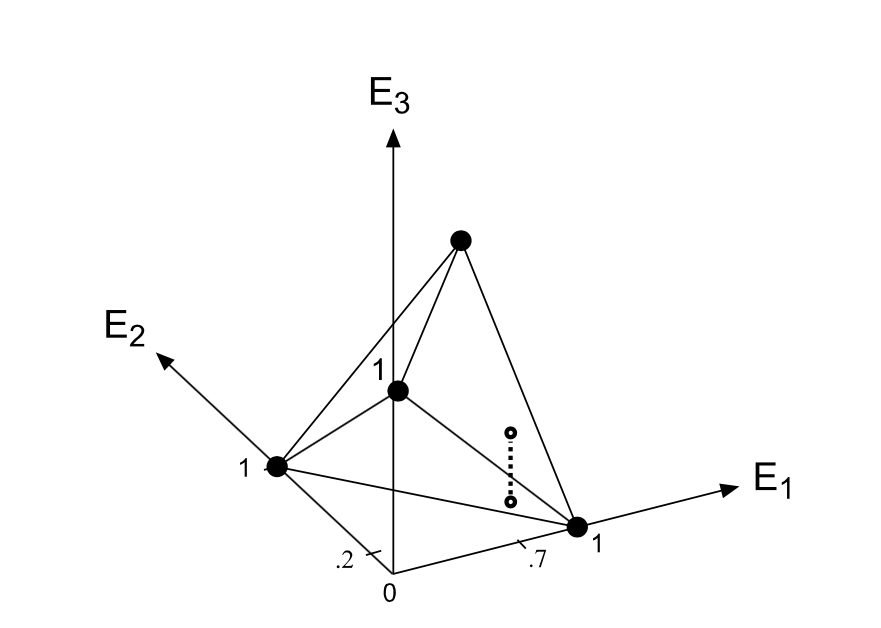}
\parbox{14cm}{\caption{\small
The convex hull of the column vectors in the realm
matrix for $(E_1, E_2,E_3)^T$.  The constraints
$P(E_1) = .7$ and $P(E_2)
= .2$ restrict the cohering assertion of $P(E_3)$ to lie within limits
specified by the endpoints of the dashed line segment touching the
boundaries of the convex hull. }}
\label{Fi:linprogdisplay1}
\end{center}
\end{figure}

Figure 2 exhibits
another interesting application of the FTP, in its more
general form as the Fundamental Theorem of Prevision.
Consider the same events composing {\bf E}$_3$, as above.  Now
suppose that the assertion conditions specified in the problem are
firstly, that the events $E_1$ and $E_2$ are regarded exchangeably, and
secondly that $P(E_1) = .7$.  Geometrically, the assertion of
exchangeability requires that the prevision vector $P(${\bf E}$_3)$ must
lie on the shaded triangular plane
displayed in Figure 2.  The points on this
plane contain all the triples of $P(${\bf E}$_3)$ possibilities
for which $P(E_1\widetilde{E}_2) = P(\widetilde{E}_1E_2)$.
Now asserting further that $P(E_1) = .7$
requires that the probability vector $P(${\bf E}$_3)$ must lie on the
line segment within this plane whose first two components equal .7.  The
resulting bounds on $P(E_3)$ are determined by the endpoints
of this line segment.

\begin{figure}[htbp]
\begin{center}
	\includegraphics[width=0.8\linewidth]{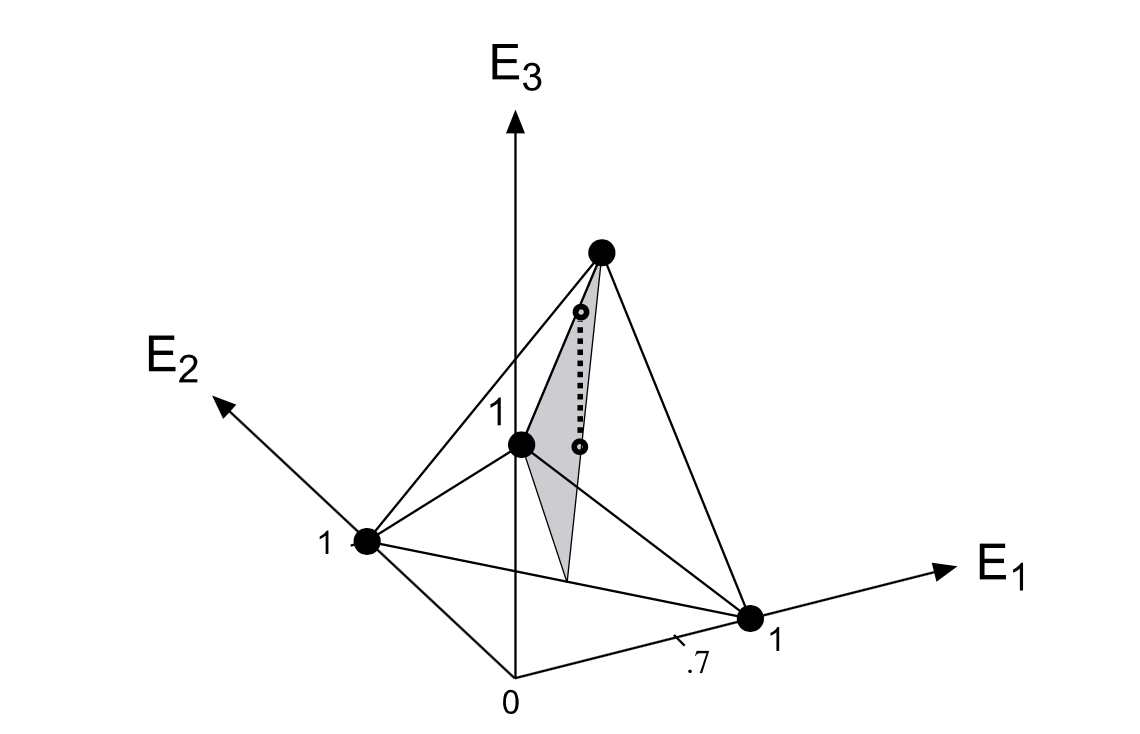}
\parbox{14cm}{\caption{\small
The convex hull of the
realm elements is identical to that displayed in Figure 1.
Asserting exchangeability of $E_1$
and $E_2$ restricts the coherent prevision vectors to those that lie on
the shaded plane.  The further assertion of $P(E_1) = .7$ constrains
the limits on a cohering assertion of $P(E_3)$ to the endpoints of the
dashed line segment.}}
\label{Fi:linprogdisplay2}
\end{center}
\end{figure} 
Computationally, the conditions of this
adjusted problem mean that the constraints on {\bf q}$_4$ change from
those specified in equation (\ref{eq:probconsteqn}) to
those in equation (\ref{eq:exmple2consteqn}).
The first row constraint on
{\bf q}$_4$ in equation (\ref{eq:exmple2consteqn})
represents the exchangeability constraint, $q_2 = q_3$;
the second row constraint represents the assertion $P(E_1) = .7\,$;  
the third row designates the summation constraint that the
components of {\bf q}$_4$ sum to 1.
\begin{equation}
P \, \left( \begin{array}{c}
\ E_1\widetilde{E}_2 - \widetilde{E}_1E_2\ \\
\ E_1\ \\
\ \ 1 \
\end{array} \right) \; \; \;  =  \; \; \;
\left( \begin{array}{c}
 \,0 \ \ -1 \ \ \ \ 1 \ \ \ \ 0 \\
\ \,0 \ \ \ \ \ \, 0 \ \ \ \ 1 \ \ \ \ 1\ \\
\ \,1 \ \ \ \ \ \, 1 \ \ \ \ 1 \ \ \ \ 1\
\end{array} \right) \ {\bf q}_4 \; \; \; = \; \; \;
 \left( \begin{array}{c}
\ \,0\ \\
\ .7\ \\
\ \,1 \
\end{array} \right)  \label{eq:exmple2consteqn}
\end{equation}
The numerical solution to the modified linear programming
problems are that the minimum coherent
value of $P(E_3)$ is .40, which corresponds to the vector
{\bf q}$_4^*(min)
\ = \ (0, .3, .3, .4)^T\,$, while the maximum value for $P(E_3)$ is
is 1.0, corresponding to {\bf q}$_4^*(max) \ = \ (.3,0,0,.7)^T$. 

\section{Framing the accuracy of median diagnosis with the FTP}

The limited information available in the asbestosis diagnosis
problem makes it natural to assess in the format provided by the
fundamental theorem of prevision.  Neither histological examination
of lung tissue nor X-ray examination by B-readers is commonly
conducted among patients who are not suspected of having asbestosis.
Moreover, histological exams are rare even among people who do have
asbestosis because of their intrusive nature.  However, Tweedie and 
Mengersen (hereafter T-M, 1999) identified two quantities about
which relevant information is available:  the frequency of positive
median diagnoses among a population of patients who present
themselves for asbestosis diagnosis via X-ray;  and the frequency
among such positive diagnoses with which the median diagnosis is
determined by a split decision.  We designate the event of a positive 
median diagnosis by $D^*$, and
the event that such a positive diagnosis arises from a split
decision by $S^*$. The realm for the four basic
events along with $D^*$ and $S^*$ is
\begin{equation}
{\bf R} \; \left( \begin{array}{c}
F\\
D_1\\
D_2\\
D_3\\
^{\_\ \_\ \_\ \_\ \_\ \_\ \_\ \_\ }\\
D^* \equiv (\sum D_{i=1}^3 \geq 2)\\
S^* \end{array} \right) \; \; \;  =  \; \; \;
\left( \begin{array}{cccccccccccccccc}
0&1&0&0&0&1&1&1&0&0&0&1&1&1&0&1\\
0&0&1&0&0&1&0&0&0&1&1&0&1&1&1&1\\
0&0&0&1&0&0&1&0&1&0&1&1&0&1&1&1\\
0&0&0&0&1&0&0&1&1&1&0&1&1&0&1&1\\
^{\_}&^{\_} & ^{\_}&^{\_} & ^{\_}&^{\_} & ^{\_}&^{\_}&^{\_}&^{\_}
&
^{\_}&^{\_} & ^{\_}&^{\_} & ^{\_}&^{\_} \\
0&0&0&0&0&0&0&0&1&1&1&1&1&1&1&1\\
0&0&0&0&0&0&0&0&1&1&1&1&1&1&0&0
\end{array} \right) \ . \ \ 
\label{eq:realm}
\end{equation} \\

The events relevant to the accuracy rates
of median decisions, $F$ and $D^*$, are both defined in terms of linear
combinations of the partition of events corresponding to
the columns of {\bf R}.  Although information is
not available to specify probabilities for each of these columns,
information that we can provide
does place restrictions on their values: \\
\noindent {\bf 1.}  These sixteen probabilities, which we
designate in vector notation by {\bf q}$_{16}$, must sum to 1.\\
\noindent {\bf 2.} Conditional exchangeability
among $D_1, D_2$ and $D_3$ given F requires that $q_6 = q_7 = q_8$
and $q_{12} = q_{13} = q_{14}$.  For the numerical values of $D_1,
D_2$ and $D_3$ in the associated columns of {\bf R} are merely
permutations of one another, and in each of these columns the value
of $F = 1$. Similarly, exchangeability conditional on $\tilde{F}$
requires that 
$q_3 = q_4 = q_5$ and $q_9 = q_{10} = q_{11}$. \\
\noindent {\bf 3.}  Information discussed by T-M 
motivates assertion values of $P(D^*) = .12$ and $P(S^*|D^*) = .42$.
Equivalently for the latter, $P(S^*D^*) = .0504$. \\
\noindent All together, these assertions amount
to eleven linear restrictions on the components of {\bf q}$_{16}$.
This
leaves five free dimensions to the specification of {\bf q}$_{16}$.  We
use programming procedures specified by the FTP to determine 
bounds on the accuracy probabilities for the median diagnosis
procedure that
cohere with these input restrictions.  \\

The analysis of T-M involved a stronger assumption, that the  
individual assessors' diagnosis decisions are conditionally
independent given both $F$ and $\tilde{F}$.  Under these assumptions, the
components of {\bf q}$_{16}$ could be determined by only three
probabilities:  $P(F), P(D_i|F)$ and $P(\widetilde{D}_i|\tilde{F})$.  Since
these are not directly available, they used the values they assessed for
$P(D^*)$ and $P(S^*D^*)$ for two of their inputs, and then required only
a third probability to complete the determination of {\bf q}$_{16}$.  The
design of their investigation was to propose a range of reasonable
possibilities  for the unknown value of $P(D_i|F)$ suggested by relevant
research literature, and to consider the reasonability of the solutions
they imply for the accuracy rates for  median decisions.
All things considered, their
ultimate comparison boiled down to the relative reasonability of 
two possible assertion values, $P(D_i|F) = .82$ and $P(D_i|F) = .90$.  \\

What we have done is append each of these two assertion 
possibilities in turn 
to the eleven restrictions on {\bf q}$_{16}$ listed above, and
compute the bounds on the median decision accuracy probabilities 
implied by the FTP.  Before comparing the results, we 
have one more set of inputs to discuss.  \\

The similar qualifications of the three B-readers
support further a partial ordering of conditional probabilities
based on their three separate diagnoses, seventeen inequalities in 
all. We shall display and interpret one array of five inequalities,
and then show why they amount to quadratic inequalities conditions
on {\bf q}$_{16}$.  A complete presentation of the 17 inequalities
is available in the TAS on-line Repository. Consider the following
row of assumed inequalities:
\begin{equation}
P(D_3|\widetilde{D_2}\widetilde{D_1}\widetilde{F}) \ \leq
\ P(D_3|\widetilde{D_1}\widetilde{F}) \ \leq \ P(D_3|\widetilde{F}) \
\leq P(D_3
|F)
\  \leq \ P(D_3|D_1F) \ \leq \ P(D_3|D_2D_1F) \ \ \ .
\label{eq:firstineqrow}
\end{equation}
To begin, the middle inequality expresses the view
that positive X-ray diagnosis of a patient with fibrosis is assessed
with higher probability than a positive diagnosis for a patient without
fibrosis.  The next inequality to
the right expresses the realization that in the
context of a patient who has fibrosis, the condition of
positive diagnosis by one B-reader motivates a greater expectation
of positive diagnosis by the next reader than would be expected without 
conditioning on the positive diagnosis by the first.  Furthermore,
a second positive diagnosis
would increase our expectation of a positive diagnosis by the third
reader even more.  This is the content of the final inequality on the
right.  The inequalities to the left of the central one express the
same structure of expectations conditioned on $\tilde{F}$ when we are
informed of negative diagnoses. \\

Now consider for example the first inequality in line
(\ref{eq:firstineqrow}): $\ \
P(D_3|\widetilde{D_2}\widetilde{D_1}\widetilde{F}) \ \leq \
P(D_3|\widetilde{D_1}\widetilde{F})$.$\ $ This is an inequality on
two conditional probabilities where the conditioning events are
different from one another. Multiplying both sides by the product
$P(\widetilde{D_2}\widetilde{D_1}\widetilde{F})
P(\widetilde{D_1}\widetilde{F})$ then yields the equivalent product
inequality
\begin{equation}
P(D_3\widetilde{D_2}\widetilde{D_1}\widetilde{F})\
P(\widetilde{D_1}\widetilde{F})
\ \ \leq
\ \
P(D_3\widetilde{D_1}\widetilde{F}) \
P(\widetilde{D_2}\widetilde{D_1}\widetilde{F
})
 \ \ .
\label{eq:equivalentineq}
\end{equation}
Both multiplicand probabilities on both sides of
(\ref{eq:equivalentineq}) are expressible as linear functions of
{\bf q}$_{16}$.  Thus, the product inequality of line
(\ref{eq:equivalentineq}) constitutes a quadratic inequality on {\bf
q}$_{16}$:
\begin{equation}
q_5 \ (q_2 + q_7 + q_8 + q_{12}) \ \ \leq \ \ (q_5 + q_9) \ (q_1 + q_5)
\ \ \ .
\end{equation}
Each of the seventeen inequality conditions inserted into our problem 
induces a 
quadratic inequality in a similar way.  These are referred to as
``further inequality conditions''  (CondExFIC) in the results of the
quadratic programming computations displayed in the next Section.

\section{Numerical results}

 Table \ref{tab:results1} contains the computational results of bounds
on accuracy probabilities both for individual radiologists and
for median decisions implied by their
coherency with the linear and quadratic conditions we have motivated.
To begin their evaluation, notice in comparing column 1 with 2 and
column 3 with 4 that the probabilities computed according to the
independence assumptions are at or near the endpoints of the
intervals allowed by the Conditional Exchangeability assumptions.  This
is particularly noticeable in the second bank of probabilities
relevant to median decisions.  This makes sense because the
independence assumption
entails that the amount of information gained by consulting an
additional radiologist (or two) is the maximum possible. Knowledge of
the diagnosis of any one radiologist
is presumed to provide no information about the
diagnosis of any other. \\

\begin{table}[ht]
\caption{BOUNDS ON PROBABILITIES based on the
assertions $P(D^*) = .12$ and
$P(S^*|D^*) = .42$, along with further assertions appropriate
to each column.  The first column of numbers 
displays the T-M probabilities based on the further assertions of
conditional independence and 
$P(D_i|F) = .82$.  The second column, labeled CondExFIC,
presumes the same three numerical probabilities, but presumes instead
conditional exchangeability along with 
the ``further inequality conditions'' described in Section 5.
Results in the next pair of
columns are based on the same presumptions as the first two,
except that $P(D_i|F)$ is specified as .90.
Presumptions for the
column headed CondExBnd are the same as for columns 2 and 4
except that only interval bounds $[.82, .90]$ are specified for
$P(D_i|F)$ and $P(\tilde{D_i}|\tilde{F})$.  The final column of
bounds, headed CondExBPlus, presumes one additional condition, 
that $P(F|D_i) \geq
.50)$.}
\label{tab:results1}
\begin{center}
\begin{tabular}{ccccccc}

{\bf Probability} & {\bf T-M} & {\bf CondExFIC} & {\bf T-M} & 
{\bf CondExFIC} & {\bf CondExBnd} & {\bf CondExBPlus}\\

 & & & & & & \\

$p = P(D_i|F)$ & $\ .82^*$  & .82$^*$ &  .90$^*$ & $\ .90^*$
& (.82\ , .90\ ) & (.82\ , .846)\\

$1-p_f = P(\widetilde{D_i}|\widetilde{F})$ & \ \ .958* & (.797, .992) &
\ .894*  & (.797, .956) & (.82\ , .90\ ) & (.898, .90\ )\\

$PV+_{ind} = P(F|D_i)$ & .734 & (.000, .932) & .466 & (.000, .657) &
(.000, .506) & (.50\ , .506) \\

$PV-_{ind} = P(\widetilde{F}|\widetilde{D_i})$ & .974 & (.973, 1.00)
& .989
& (.988, 1.00) & (.975, 1.00) & (.975, .981) \\

\_\_\_\_\_\_\_\_\_\_\_ & \_\_\_\_\_ & \_\_\_\_\_\_\_ & \_\_\_\_\_ &
\_\_\_\_\_\_\_ & \_\_\_\_\_ & \_\_\_\_\_\_\_  \\

$P(D^*|F)$ & .914 & (.820, .915) & .972 & (.900, .972) & (.82\ , .972) &
(.852, .898)\\

$P(\widetilde{D^*}|\widetilde{F})$ & .995 & (.880, .995) & .968
& (.880, .969) & (.880, .972) & (.969, .972) \\

$PV+ = P(F|D^*)$ & \ .961* & (.000, .961) & $\ .761*$ & (.000, .762)
& (.000, .793) & (.772, .793)\\

$PV- = P(\widetilde{F}|\widetilde{D^*})$ & \ .987* & (.979, 1.00) &
\ .997* & (.990, 1.00) & (.979, 1.00) & (.981, .988)\\

\_\_\_\_\_\_\_\_\_\_\_ & \_\_\_\_\_ & \_\_\_\_\_\_\_ & \_\_\_\_\_ &
\_\_\_\_\_\_\_ & \_\_\_\_\_ & \_\_\_\_\_\_\_  \\

$P(F)$ & .126 & (.000, .127) & .094 & (.000, .094) & (.000, .111) &
(.105, .111)\\

\_\_\_\_\_\_\_\_\_\_\_ & \_\_\_\_\_ & \_\_\_\_\_\_\_ & \_\_\_\_\_ &
\_\_\_\_\_\_\_ & \_\_\_\_\_ & \_\_\_\_\_\_\_  \\

\end{tabular}
\end{center}
\end{table}

Some of the targeted probabilities of
interest are bounded rather tightly by conditional exchangeability and
the further inequality conditions, while others are not.
The CondExFIC columns 2 and 4 (entailing $p = .82$ and $p = .90$)
show fairly tight
and nearly equivalent ranges for cohering assertions of the negative
predictive values $PV-_{ind}$
and the median $PV-$;  and nearly
equivalent and broader but still useable intervals for both individual
and median specificities 
$P(\widetilde{D_i}|\widetilde{F})$ and
$P(\widetilde{D^*}|\widetilde{F})$.
However, the ranges for positive predictive values $P(F|D_i)$ and
$P(F|D^*)$  are very broad.
Upper bounds differ by .275 and .199 when individual
and median diagnoses are compared at the specifications of $p = .82$ and
$p = .90$.\\

One of the uses of bounding results for a wide array of relevant
probabilities is to identify further conditions that would help to
narrow our focus on important probabilities of interest.  One such
condition arises from the allowable bounds on the individual
radiologists' positive predictive value, $PV+_{ind}$.  Both the
CondExFIC columns show probability intervals with a lower
bound of zero would cohere with the assertions we have input to
the problem.  However, even a minimal respect for the value of radiological
assessment would impose a restriction that the positive
predictive value of a radiologist's diagnosis exceeds .5.   
Would you rather bet on $F$ if a B-reader made a positive diagnosis, or
would you rather bet on $\tilde{F}$?  If you would rather bet on $F$,
you would want 
to bound $PV+_{ind}$ above .5.  We have conferred with a very
experienced clinical and research oncologist who confirmed that this
would be a minimal requirement of assessment probabilities, agreed by
virtually every knowledgeable oncologist.
It is worth noticing in this regard that the value of
$PV+_{ind}$ implied in the T-M(p=.9) analysis is .466, based on their
conditional independence assumption.  Yet the upper bound on $PV+_{ind}$
that coheres with this sensitivity value is as high as .657 if only
conditional exchangeability is presumed.  Assuming
only conditional exchangeability would free the assessment of all
accuracy probabilities to allow sensible ranges. \\

T-M evaluated their array of implied probabilities in a limited way
by questioning the plausibility ``that the true positive rate [of an
individual diagnosis] should be as low as .82 reported in Kipen et
al (1987) since this implies the true negative rate is extremely
high'' (T-M, 1999, p. 237).  They were alluding to the implied
probability $1-p_f = .958$ in the T-M($p=.82$) column.  
However, the lower bound allowable under exchangeability is as low
as .797 in this case (the same as the lower bound implied by their
suggested choice of $p = .90$).  In their discussion of the
situation, T-M suggest that perhaps sensitivity and specificity
might be presumed to be about equal, a feature that further
motivated the choice of $p = .9$ in their subsequent
analysis.  \\

We have followed this thread of T-M's suggestion in the following
way. The fifth column of numerical results, headed CondExBnd, again
presumes the probabilities $P(D^*) = .12$ and $P(S^*|D^*) = .42$
along with the conditional exchangeability of the individual
radiologists' $D_i$ given $F$ and $\tilde{F}$.  Rather than
specifying an exact probability for $P(D_i|F)$, we merely assert a
bound that both probabilities $P(D_i|F)$ and
$P(\widetilde{D_i}|\tilde{F})$ must lie within the interval $[.82,
.90]$ to represent our uncertainty.  
As might be expected, most of the probability bounds
appearing in this column virtually cover the intersection of the
bounds specified in the CondExFIC columns with $p = .82$ and $p =
.90$.  The only really noticeable exceptions occur for the
predictive 
probabilities $P(F|D_i)$ and $P(F|D^*)$.  The former is now bounded
within the interval $(.000, .506)$ and the latter within $(.000,
.793)$.  In particular, the interval restrictions on $P(D_i|F)$ and
$P(\widetilde{D_i}|\tilde{F})$ rule out the higher end of the range
on these
probabilities allowed when $p = .82$.  \\

In the sixth and final column we add the further restriction that the
positive predictive value for an individual B-reader should at least
exceed .50.  The consequences of adding this reasonable assumption are
rather severe and illuminating.
In the first place, the sensitivity probability $P(D_i|F)$
is now bounded well away from .90, rather within the fairly tight interval
$(.82, .846)$.  Moreover, the specificity probability for an individual
B-reader, $P(\widetilde{D_i}|\tilde{F})$, is now bounded tightly as well,
within the interval $(.898, .90)$.  This rules out the higher
probabilities that the specification of $p = .82$ allows and
narrows the interval virtually to equal the value that had been
preferred by T-M, motivating their choice of $p = .9$ in the context of
presumed independence.  At the same time, coherency forces the value of
$p$ much closer to .82 than to .9.  In fact, all of the accuracy
probabilities displayed in the final column compare quite reasonably
with the values proposed in the T-M($p=.9$) column {\it except for the
values of} $p=.9$ and $PV+_{ind} = .466$.  
Moreover, the values they
assessed for median $PV-$, which might well have
been regarded as too high, are tempered a bit, while their unduly low
positive predictive values are boosted somewhat.  In sum, the assertion
of conditional exchangeability supports an assertion of $P(D_i|F)$
around $.82$ rather than $.9$.  The realism of assumptions allowed by de
Finetti's FTP are critical to understanding this result. \\

A complete discussion of many related computations is beyond the 
scope of this article.  More details appear in the research report of 
Capotorti, Lad and Sanfilippo (2003).  This includes a
commentary on the GAMS computing software which was used for our
computational results.  See Brooke et al. (2003).
%

\section*{References}

\noindent {\bf Brooke, A., Kendrick, D., Meeraus, A. and Raman, R.} (2003)
{\it GAMS: a User's Guide}, Washington, D.C.:  GAMS Development Corp.
\\

\noindent {\bf Capotorti, A., Lad, F., and Sanfilippo, G.} (2003)
Reassessing accuracy rates of median decision procedures,
University of Canterbury Department of Mathematics and Statistics
Research Report, DMS 2003/21, 
www.math.canterbury.ac.nz/php/research/reports. \\

\noindent {\bf de Finetti, B.} (1937) La pr\'{e}vision, ses
lois logiques, ses sources subjectives, {\it Ann. Inst. H.
Poincar$\acute{e}$}, {\bf 7}, pp. 1-68.  H. Kyburg (tr.) Foresight,
its logical laws, its subjective sources, in H. Kyburg and H.
Smokler (eds.) {\it Studies in
Subective Probability}, second edition, 1980, New York: Krieger. \\

\noindent {\bf de Finetti, B.} {\it Theory of Probability: a critical
introductory treatment} (1974,1975)
 2 volumes,
A.F.M. Smith and A. Machi (trs.), New York: Wiley. Translation of {\it
Teoria della probabilita: \ sintesi introduttiva con appendice critica}
(1970) Torino:  Einaudi.  \\

\noindent {\bf Heath, D. and Sudderth, W.} (1976) De Finetti's
theorem for exchangeable random variables, {\it Amer. Stat.}, {\bf
30}, 188-189.  \\

\noindent {\bf Lad, F.} (1996) {\it Operational Subjective
Statistical Methods: a mathematical, philosophical, and historical
introduction}, New York: John Wiley. \\

\noindent {\bf Kipen, H.M., Lilis, R., Suzuki, Y., Valciukas, J.A.,
and Selikoff, I.J.} (1987) Pulmonary fibrosis in asbestos insulation
workers with lung cancer:  a radiological and histopathological
evaluation, {\it Brit. Journ. Indust. Med.}, {\bf
44}, 96-100.  \\

\noindent {\bf Tweedie, R. and Mengersen, K.} (1999)  Calculating
accuracy rates from multiple assessors with limited information,
{\it Amer. Stat.}, {\bf 53}, 233-238. \\

\pagebreak

\begin{center}

\LARGE

{\bf Materials 
Associated with this
Article}\\

\end{center}

\normalsize

We have prepared three sections of extensions to sections of our article
that are necessary for a complete statement of precisely how we have
made the computations.  For much more extensive discussion of all
results and even more results, please consult the research report of
Capotorti, Lad and Sanfilippo (2003) which is referenced in the article
itself.  \\

\section*{Appendix 1: Extensions to the four-step computational \\
procedure appropriate to
the most general form of the FTP}

We generalize the formal introduction to the FTP here 
by merely stating some
extensions to the four step procedure that are prescribed in its most
general form.  For details you may consult a reference such as
Lad (1996).  Any number of previsions or prevision inequalities 
may be asserted as
conditions for the theorem, and the number of constituents in the
relevant partition will depend on the number of quantities involved
and on the extent of the logical
relations among them.\\
\noindent {\bf a.}   References to ``events'' in the
procedure can be replaced by ``quantities'' and corresponding references
to probabilities can be replaced by the unifying concept of
``prevision.'' \\
\noindent {\bf b.}  Conditional previsions can be included among the
assertions given in the suppositions of the theorem, and these will
imply linear constraints on the relevant vector {\bf q}.\\
\noindent {\bf c.}  Prevision orderings (inequalities) and intervals
for previsions
may be included among the conditions without affecting the linearity
structure. \\
\noindent {\bf d.}  Orderings of conditional previsions that involve
different conditioning events are allowable too.  However, the
constraints these imply on {\bf q} would be quadratic rather than
linear.  (We briefly discussed why this is the case 
in the asbestosis diagnosis problem.)\\
\noindent {\bf e.}  The object of the enquiry assessed in the theorem
can be a conditional prevision too, without affecting the linear
structure of the objective function.  However, an appropriate
transformation of the problem is required to change the ostensibly
rational (fractional) objective function into a linear function.\\

\section*{Appendix 2:  Twelve more 
inequalities involving Conditional \\ Probabilities that were 
assumed in the computational results headed ``CondexFIC''}

The next four rows of inequalities shall be presented in pairs without
discussion.  Their motivation is similar to the examples outlined in the
text of the article.  For now, you are
left to interpret them and to assert their reasonability for this
analysis yourself.  You will find them discussed in the technical report
mentioned above.
\begin{eqnarray}
P(D_2|\widetilde{D_1}\widetilde{F}) \ &  \leq & \
P(D_2|\widetilde{D_1}F) \
\ \leq \  \ P(D_2|F) \ \ \ \ \   {\rm and} \nonumber \\
P(D_2|\widetilde{F}) \ & \leq & \ P(D_2|D_1\widetilde{F}) \ \ \leq \ \
P(D_2|D_1
F) \ \
\ \ \ \ \ \ \ \ \ \ \ ; \ and
\label{eq:secondineqrows}
\end{eqnarray}
\begin{eqnarray}
P(D_3|\widetilde{D_2}\widetilde{D_1}F) \ &  \leq & \
P(D_3|\widetilde{D_1}F) \
\ \leq \ \ P(D_3|\widetilde{D_1}D_2F) \ \ \ \ \   {\rm and} \nonumber
\\
P(D_3|\widetilde{D_2}D_1\widetilde{F}) \ & \leq & \
P(D_3|D_1\widetilde{F}) \
\ \leq \  \ P(D_3|D_2D_1\widetilde{F}) \ \ \ .
\label{eq:thirdineqrows}
\end{eqnarray}

The final row of inequalities is centered by a numerical bound.
\begin{equation}
P(D_3|\widetilde{D_1}\widetilde{F}) \ \leq \
P(D_3|\widetilde{D_1}D_2\widetilde{
F})
\ \leq  .5 \ \leq \  P(D_3|\widetilde{D_2}D_1F) \ \leq \ P(D_3|D_1F) \
\ \ .
\label{eq:fourthineq}
\end{equation}

\noindent  In considering the inequalities around .5, think of a
question such as this:  if you found a patient who suffers from 
asbestosis (so $F=1$ even if this is unbeknownst to you) 
and you learned that two B-readers made a positive
and negative diagnosis based on an X-ray, would you rather bet \$1 on
the third B-reader making a positive diagnosis or would you rather bet
on your flipping a head with a coin in your pocket?  If you would rather
bet on a positive diagnosis by the third radiologist, then you
conditional probability $ P(D_3|\widetilde{D_2}D_1F) \ \leq \
P(D_3|D_1F)$ is bounded above $.5$.\\

\pagebreak

\section*{Appendix 3:  More Numerical bound results} 

{\bf To the editor:}  For now, we are merely appending the 
bounds on the differences in accuracy
probabilities for individual and median decisions that have been deleted
from the previous submitted edition.  These shall be supplemented
further with bounds on a complete list of relevant probabilities and
conditional probabilities before the article goes to press.  We are 
currently in the process of extending the computations and constructing
the table.\\

The table below appends to Table 1 shown in the text the bounds on the
difference in accuracy probabilities between individual diagnosis
decisions and median decisions.  The largest gains in accuracy occur in
the value of positive predictive values, though gains in sensitivity and
specificity of the diagnosis are also of a size that is recognizable.
Negative predictive values are already understood to be quite large for
the accuracy of individual diagnoses, at least on the order of $.975$.  
Thus, absolute gains are not large. Again, as interpreted in the
article, the most interesting and sensible of the columns of results is
the final one, headed CondExBPlus.  \\

\begin{table}[ht]
\caption{FURTHER BOUNDS ON DIFFERENCES IN ACCURACY PROBABILITIES between
individual and median decisions, continuing the assumptions described in
the caption to Table 1 of bounds that appear in the article.}
\label{tab:results2}
\begin{center}
\begin{tabular}{ccccccc}

Probability & T-M & CondExFIC & T-M & CondExFIC & CondExBnd &
CondExBPlus\\

 & & & & & & \\

$p = P(D_i|F)$ & $\ .82^*$  & .82$^*$ &  .90$^*$ & $\ .90^*$
& (.82\ , .90\ ) & (.82\ , .846)\\

$1-p_f = P(\widetilde{D_i}|\widetilde{F})$ & \ \ .958* & (.797, .992) &
\ .894*  & (.797, .956) & (.82\ , .90\ ) & (.898, .90\ )\\

$PV+_{ind} = P(F|D_i)$ & .734 & (.000, .932) & .466 & (.000, .657) &
(.000, .506) & (.50\ , .506) \\

$PV-_{ind} = P(\widetilde{F}|\widetilde{D_i})$ & .974 & (.973, 1.00)
& .989
& (.988, 1.00) & (.975, 1.00) & (.975, .981) \\

\_\_\_\_\_\_\_\_\_\_\_ & \_\_\_\_\_ & \_\_\_\_\_\_\_ & \_\_\_\_\_ &
\_\_\_\_\_\_\_ & \_\_\_\_\_ & \_\_\_\_\_\_\_  \\

$P(D^*|F)$ & .914 & (.820, .915) & .972 & (.900, .972) & (.82\ , .972) &
(.852, .898)\\

$P(\widetilde{D^*}|\widetilde{F})$ & .995 & (.880, .995) & .968
& (.880, .969) & (.880, .972) & (.969, .972) \\

$PV+ = P(F|D^*)$ & \ .961* & (.000, .961) & $\ .761*$ & (.000, .762)
& (.000, .793) & (.772, .793)\\

$PV- = P(\widetilde{F}|\widetilde{D^*})$ & \ .987* & (.979, 1.00) &
\ .997* & (.990, 1.00) & (.979, 1.00) & (.981, .988)\\

\_\_\_\_\_\_\_\_\_\_\_ & \_\_\_\_\_ & \_\_\_\_\_\_\_ & \_\_\_\_\_ &
\_\_\_\_\_\_\_ & \_\_\_\_\_ & \_\_\_\_\_\_\_  \\

$P(F)$ & .126 & (.000, .127) & .094 & (.000, .094) & (.000, .111) &
(.105, .111)\\

\_\_\_\_\_\_\_\_\_\_\_ & \_\_\_\_\_ & \_\_\_\_\_\_\_ & \_\_\_\_\_ &
\_\_\_\_\_\_\_ & \_\_\_\_\_ & \_\_\_\_\_\_\_  \\

$P(D^*|F)-P(D_i|F)$ & .094 & (.000, .095) & .072 & (.000, .072)
& (.000, .095) & (.032, .055)\\

$P(\widetilde{D^*}|\tilde{F})-P(\widetilde{D_i}|\tilde{F})$ & .037 &
(.000, .088) & .075 & (.000, .088) & (.000, .088) & (.069, .073) \\

$P(F|D^*)-P(F|D_i)$ & .227 & (.000, .299) & $.294$ & (.000, .295)
& (.000, .299) & (.272, .293)\\

$P(\tilde{F}|\widetilde{D^*})-P(\tilde{F}|\widetilde{D_i})$ & .014 &
(.000, .015) & .008 & (.000, .009) & (.000, .012) & (.005, .009)\\

\_\_\_\_\_\_\_\_\_\_\_ & \_\_\_\_\_ & \_\_\_\_\_\_\_ & \_\_\_\_\_ &
\_\_\_\_\_\_\_ & \_\_\_\_\_ & \_\_\_\_\_\_\_  \\

\end{tabular}
\end{center}
\end{table}

\end{document}